\documentclass[fleqn]{2020SCGE} %, fleqn
\usepackage{siunitx}
\usepackage{hyperref}
\usepackage{booktabs}
\usepackage{lineno}
\usepackage{color}
\usepackage{verbatim}
\usepackage{float}

\usepackage{graphicx}
\usepackage{amsmath}
\usepackage{amssymb}	% Extra maths symbols
\usepackage{diagbox}
\usepackage{threeparttable}

\usepackage{ulem}
\usepackage[numbers]{natbib}
\usepackage{longtable, threeparttablex, booktabs, url}
\usepackage{multirow} % 支持合并单元格
\usepackage{array} % 改进表格功能
\usepackage{makecell} % 支持表格中的自动换行
\usepackage{colortbl}
\usepackage{xcolor}
\usepackage{tabularx} % 自适应表格宽度
\usepackage[marginal]{footmisc}
\renewcommand{\thefootnote}{}
\usepackage{color}
\begin{document}
%\linenumbers 
\ensubject{subject}
%%%%%%%%%%%%%%%%%%%%%%%%%%%%%%%%%%%%%%%%%%%%%%%%%%%%%%%
%%% Authors do not modify the information below
%Letter to the Editor
\ArticleType{Article}
\SpecialTopic{SPECIAL TOPIC: }
\Year{2025}
\Month{xxx}
\Vol{xx}
\No{x}
\DOI{xx}
\ArtNo{000000}
\ReceiveDate{xx}
\AcceptDate{xx}
%%% Author information for page head. 
\AuthorMark{Renzhi Su}

%%% Authors for citation. 
\AuthorCitation{ Renzhi Su, et al.}
%\OnlineDate{January 1, 2016}
%%%%%%%%%%%%%%%%%%%%%%%%%%%%%%%%%%%%%%%%%%%%%%%%%%%%%%%
\title{A stringent constraint on the fractional change of proton g-factor}

%%% Corresponding author: 
%%%   \author[number]{Full name}{{email@xxx.com}}
%%% General author: 
%%%   \author[number]{Full name}{}
\footnote{$\ast$These authors contribute equally to this work.}

\author[1,2]{Renzhi Su\textsuperscript{$\ast$}\footnote{Email: rzsu.astro@gmail.com}}{}
\author[3]{Stephen, J., Curran\textsuperscript{$\ast$}}{}
\author[4]{Jeremy Darling}{}
\author[1]{Minfeng Gu}{}
\author[5,6]{J. N. H. S. Aditya}{}
\author[7]{\\Ningyu Tang}{}
\author[8,9]{Di Li}{}
\author[9]{Zheng Zheng}{}
%%%%

\address[1]{Shanghai Astronomical Observatory, Chinese Academy of Sciences, 80 Nandan Road, Shanghai 200030, China}
\address[2]{Research Center for Astronomical Computing, Zhejiang Laboratory, Hangzhou 311100, China}
\address[3]{School of Chemical and Physical Sciences, Victoria University of Wellington, PO Box 600, Wellington 6140, New Zealand}
\address[4]{Centre for Astrophysics and Space Astronomy, Department of Astrophysical and Planetary Sciences, University of Colorado, 389 UCB, Boulder, CO 80309-0389, USA}
\address[5]{Sydney Institute for Astronomy, School of Physics A28, University of Sydney, NSW 2006, Australia}
\address[6]{ARC Centre of Excellence for All Sky Astrophysics in 3 Dimensions (ASTRO 3D), Sydney, Australia}
\address[7]{Department of Physics, Anhui Normal University, Wuhu, Anhui 241002, People’s Republic of China}
\address[8]{New Cornerstone Science Laboratory, Department of Astronomy, Tsinghua University, Beijing 100084, China}
\address[9]{National Astronomical Observatories, Chinese Academy of Sciences, Beijing 100012, China}

\abstract{
We report a constraint on the cosmological variation of the proton g-factor, $g_p$. By comparing the measured redshifts between \mbox{H\,{\sc i}} 21 cm and OH 18 cm lines observed with the newly commissioned Five-hundred-meter Aperture Spherical radio Telescope (FAST) toward PKS 1413+135 at $z$ = 0.24671, we obtain $\Delta g_{p}/g_{p} = (-4.3\pm2.5)\times10^{-5}$, which is more than two orders of magnitude more sensitive than previous constraints. In addition, we obtain sensitive constraints of  $\Delta (\mu\alpha^{2})/(\mu\alpha^{2}) = (2.0\pm1.2)\times10^{-5}$ and $\Delta (\mu\alpha^{2}g_{p}^{0.64})/(\mu\alpha^{2}g_{p}^{0.64}) = (-4.7\pm1.9)\times10^{-6}$.

}

\PACS{......}
\maketitle

%%%%%%%%%%%%%%%%%%%%%%%%%%%%%%%%%%%%%%%%%%%%%%%%%%%%%%%
%%% The main text. 
%\twocolumn\onecolumn
%%%%%%%%%%%%%%%%%%%%%%%%%%%%%%%%%%%%%%%%%%%%%%%%%%%%%%%

\section{Introduction} \label{sec:intro}
The gyro-magnetic ratio of proton ($g_p$), as a fundamental physical constant, is predicted to exhibit temporal variations in theoretical frameworks attempting to unify the Standard Model of Particle Physics with General Relativity \citep[e.g.][]{marciano1984,damour1994,li1998}. Currently, only three observational constraints on $g_p$ have been obtained toward B0218+357 at $z$ = 0.6846, M31, and PKS 1830-211 at $z$ = 0.8858 \citep{chengalur2003,su2025,su2025_2}. However, these constraints are not stringent with a precision of $\sim10^{-3}$. There are no other tests of the variation of $g_p$ although various terrestrial experiments and astronomical observations have been carried out \citep[e.g.][]{darling2004,rosenband2008,murphy2008,curran2011,king2012,onegin2012,bagdonaite2013,kanekar2018,filzinger2023,jiang2024,wei2024}, which have constrained the variations of fine-structure constant, $\alpha$, proton-electron mass ratio, $\mu$, and combinations of $\alpha$, $\mu$ and $g_p$ to precisions of $\leq$ $10^{-5}$, see a recent review by \cite{uzan2024}.   

Here we report a constraint on the cosmological variation of $g_p$ utilizing \mbox{H\,{\sc i}} 21 cm and OH 18 cm lines  observed with the newly commissioned Five-hundred-meter Aperture Spherical radio Telescope \citep[FAST;][]{nan2011,li2018} toward PKS 1413+135. 

PKS 1413+135 is a BL Lac object towards which an optical spiral galaxy was observed at z = 0.247 \citep{perlman2002}.  Various radio absorption lines, arising from the spiral galaxy, have been detected towards the radio source, including atomic \mbox{H\,{\sc i}} (21 cm) \citep{carilli1992}, conjugate OH 18 cm satellite lines \citep[][]{darling2004,kanekar2010_2,kanekar2018,combes2023}\footnote{When the pumping is dominated by the intraladder 119 $\mu$m transition, the OH 18 cm satellite lines will appear as stimulated absorption (1612 MHz) and emission (1720 MHz) with the same strength, showing conjugate behavior while the OH 18 cm main lines are suppressed, exhibiting no absorption/emission.} and millimeter band molecular (CO, HNC and $\rm HCO^{+}$) lines \citep{wiklind1994,wiklind1997}. The latter were only seen towards the radio core where there is sufficient high frequency continuum (43 GHz) flux when the H\,{\sc i} absorption was detected towards the eastern jet \cite{carilli2000,perlman2002}. The most sensitive observation of the OH satellite lines has provided a constraint of $[\Delta Y/Y]=(-1.0\pm1.3)\times10^{-6}$, where $Y\equiv \mu\alpha^{2}g_{p}^{0.54}$ \citep{kanekar2018}, although $g_{p}$ was implicitly assumed to be constant which we do not assume here. 

\section{Observations~and~data~reduction}
The observations were carried out on 1st September 2022 with On-Off mode. The observations utilized the FAST L-band 19-beam receiver covering 1.00--1.50 GHz, over 65\,536 channels, which gives a beam size of $\sim$ 3 arcmins. The large bandwidth allows us to observe  the redshifted \mbox{H\,{\sc i}} and four OH lines simultaneously, while retaining a high spectral resolution. The On and Off time was set to 300 s, and 10 On-Off cycles were used with all the 19 beams. To optimize the observing time, we followed the same strategy as \citep{zheng2020,su2023} -- selecting a special Off point for beam 01, so that when the beam was off the source, beam 14 that is about 11.5 arcmins left to beam 01 was on the source. Therefore, both beam 01 and beam 14 were recording data, giving a total on-source time of 6000 s (100 minutes). The data were recorded every 0.1 s, and a noise diode, $\approx12.5$~K, were injected for 3 $\times$ 90 s to calibrate the flux.

We individually reduced the XX and YY polarization data from each beam and each On-Off cycle. Therefore, each line has 40 On-Off spectra. We first used the noise diode and the gain curves to calibrate the spectral flux. For each On-Off cycle, the Off-target spectrum was then subtracted from the On-target spectrum to calibrate the bandpass. Following this, a combination of a third-degree polynomial and a sine functions was fitted to line-free channels to subtract the continuum. 

All four OH 18 cm lines are free of RFI, although for eight (out of 40) of the OH 1720 spectra we could not satisfactorily fit the continuum and so these were not used. Intermittent weak RFI affected four \mbox{H\,{\sc i}} spectra, which were also removed from the analysis. A Kolmogorov-Smirnov test on the line-free channels of the remaining spectra showed the noise to be Gaussian and so these were retained. Finally, the spectra for each line were combined, weighted based on their measured root-mean-square (RMS) noise, to produce the final spectra for the individual transitions. Since the OH main lines have relatively low signal-noise-ratio, they were resampled to a frequency resolution of $\Delta\nu= 29$~kHz ($\Delta v = 6.3$~km~s$^{-1}$), using the Python-based $SpectRes$ \citep{carnall2017}.

\section{Spectra~and~redshift~measurements}
\begin{figure}[ht]
\centering
\includegraphics[width=\textwidth]{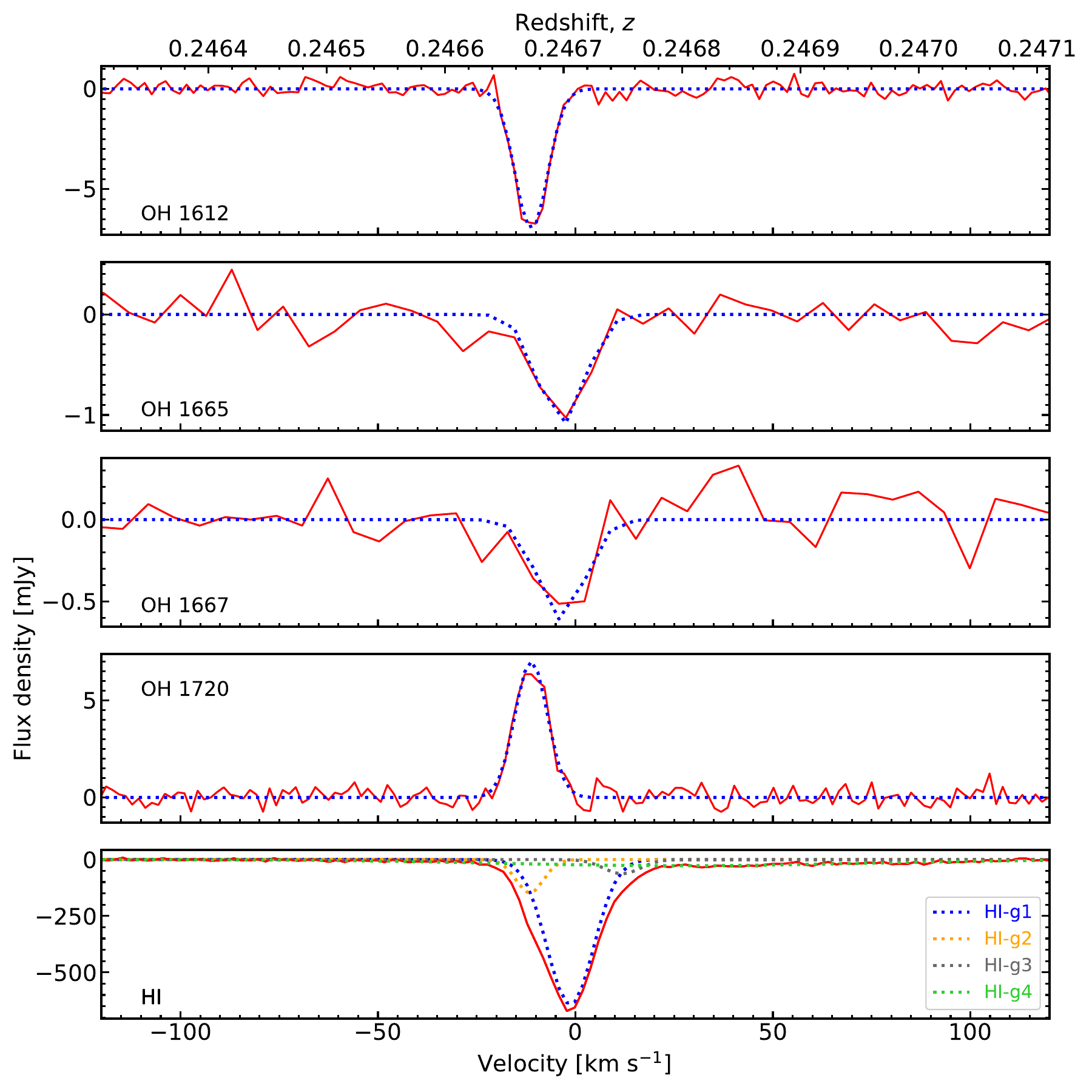}
\caption{The FAST \mbox{H\,{\sc i}} and OH spectra in the source rest frame with respect to $z$ = 0.24671. The dotted lines show the Gaussian fits. While the OH lines are fitted with single Gaussian profile, the  \mbox{H\,{\sc i}} line is best fit with four, labeled as  \mbox{H\,{\sc i}}-g1,  \mbox{H\,{\sc i}}-g2,  \mbox{H\,{\sc i}}-g3 and  \mbox{H\,{\sc i}}-g4 with decreasing peak absorption.
}\label{fig:figure1}
\end{figure}

\begin{figure}[ht]
\centering
\includegraphics[width=\textwidth]{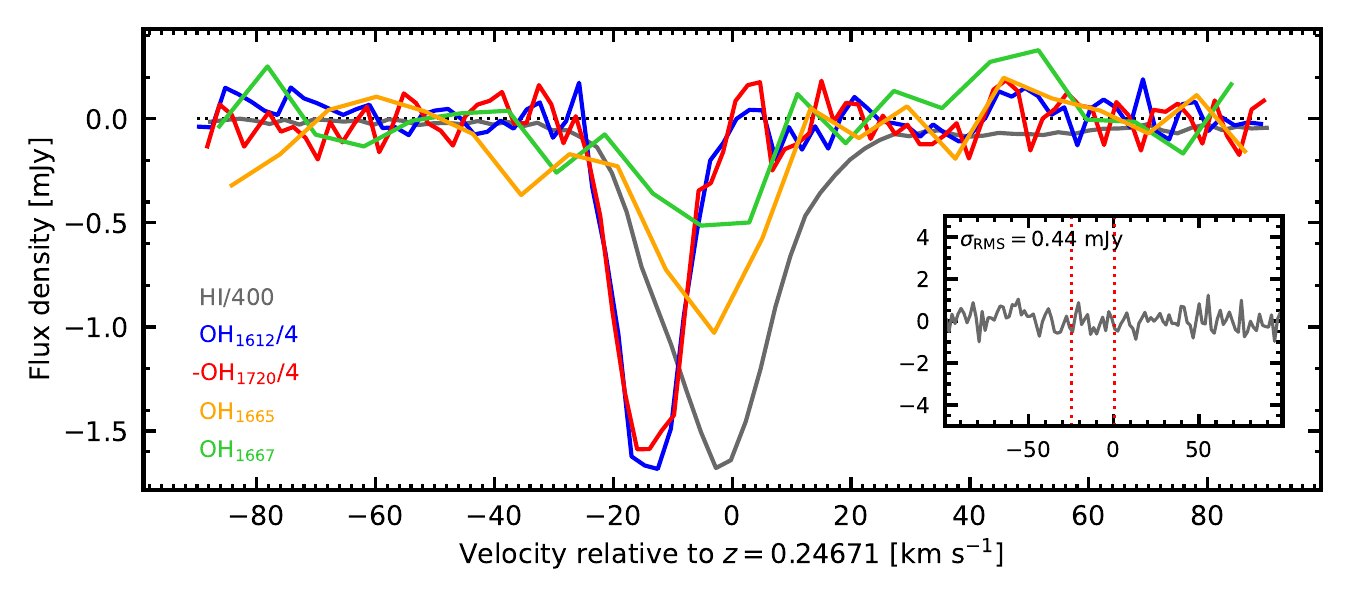}
\caption{Overlay of the  \mbox{H\,{\sc i}} and OH spectra shifted to the source rest frame. The strength of 1612, 1720 and \mbox{H\,{\sc i}} lines are divided by a factor of 4, -4 and 400, respectively. The inset shows the sum of the 1612 and 1720 lines, with the red vertical lines marking the range of the absorption which is 
consistent with the noise, as expected from the conjugate nature.}\label{fig:figure2}
\end{figure}

We show the spectra in Figure~\ref{fig:figure1} and \ref{fig:figure2}.  The spectral properties are listed in Table \ref{tab:obs_results}. We detected OH 18 cm main lines\footnote{Note that the OH 1667 line was reported to be detected toward PKS 1413+135 more than twenty years ago with Giant Metrewave Radio Telescope (GMRT) observations\citep{kanekar2002}, which was however not supported by the higher sensitive Green Bank Telescope (GBT) observations \citep{darling2004}.} toward PKS 1413+135 and they have the same velocity width. The OH 18 cm satellite lines still remain conjugate (Figure \ref{fig:figure2}), compared to previous observations \citep{kanekar2018}. 

We fitted the spectra with Gaussian functions. The \mbox{H\,{\sc i}} line has a very high signal-noise-ratio (SNR), with the peak absorption having SNR $\approx$ 1640. The high SNR reveals many subtle structures beyond the scope of our phenomenological model. The \mbox{H\,{\sc i}} line of PKS 1413+135 was observed with GBT telescope as well, which was well fitted with four components \citep{darling2004}. In principle, we can directly use the GBT spectrum to conduct our studies. However, using our FAST \mbox{H\,{\sc i}} spectrum can avoid some systematic uncertainties. Note that in our \mbox{H\,{\sc i}} line fitting practice we consistently identify a component matching the deep absorption structure whose full width at half-maximum (FWHM) is consistent with the Gaussian functions fitted to the OH 1665 and 1667 MHz lines, which indicates that these lines arise from the same gas cloud and hence we can compare their observed redshifts to measure the evolution of fundamental constants. We further checked the GBT spectrum and found this component matches the \mbox{H\,{\sc i}} 1 component in the GBT spectrum \citep{darling2004}. Therefore, we finally fitted our \mbox{H\,{\sc i}} and OH spectra following the scenario below.

\begin{itemize}
\item We deteriorated the sensitivity by manually injecting a noise of 4 mJy into the \mbox{H\,{\sc i}} spectrum, which effectively downweights the overconstrained regions while preserving the overall profile shape. This method has been used to fit the \mbox{H\,{\sc i}} spectrum of M31 \citep{su2025}. After injecting the 4 mJy noise, we found that our \mbox{H\,{\sc i}} spectrum can be fitted with four components that are consistent with the components fitted to the GBT \mbox{H\,{\sc i}} line, which demonstrates the feasibility and reliability of our method of injecting extra noise.

\item Four Gaussian components, named as \mbox{H\,{\sc i}}-g1, \mbox{H\,{\sc i}}-g2, \mbox{H\,{\sc i}}-g3 and \mbox{H\,{\sc i}}-g4, were fitted to our \mbox{H\,{\sc i}} spectrum while each OH spectrum was fitted with one Gaussian component. 

\item The FWHMs of \mbox{H\,{\sc i}}-g1, OH 1665, and OH 1667 MHz lines were tied to be the same. Here, the \mbox{H\,{\sc i}}-g1 was used to represent the deep \mbox{H\,{\sc i}} absorption structure.

\item Given the conjugate nature of the OH satellite lines, their FWHMs were tied to be the same and their amplitudes were set to be opposite.

\item To obtain the statistical uncertainties, we conducted a Monte Carlo simulation. We repeated the fits for 10,000 times. Each time, we added Gaussian random noises,
characterized by the noise values of \mbox{H\,{\sc i}} and OH spectra, to the
best fitted models of \mbox{H\,{\sc i}} and OH lines, respectively. We used the averages as the best fitted parameters and the standard deviations as associated uncertainties. 

\end{itemize}

The best fitted parameters and their associated uncertainties are listed in Table \ref{tab:obs_results}. The simultaneous fit of \mbox{H\,{\sc i}} and OH main lines gave a $\chi^2_{\text{red}}=1.04\pm0.10$ while the simultaneous fit of OH satellite lines gave a $\chi^2_{\text{red}}=1.02\pm0.06$, reflecting a good fit. 

When measuring the redshifts, the laboratory frequencies, $\nu_{\rm \mbox{H\,{\sc i}}}= 1420.405751768(2)$ MHz, $\nu_{1612}$ = 1612.230825(15) MHz, $\nu_{1665}$ = 1665.401803(12) MHz, $\nu_{1667}$ = 1667.358996(4) MHz, $\nu_{1720}$ = 1720.529887(10) MHz, were used \citep{hellwig1970,hudson2006,lev2006}.

\begin{sidewaystable}[h]
%\begin{table}[h]
\caption{The measured properties of absorption lines towards PKS\,1413+135. The columns are the transition, the peak absorption flux density, FWHM, velocity with respect to $z$ = 0.24671, statistical uncertainties of the velocity,  systematic uncertainties of the velocity, measured redshift, spectral resolution, and the RMS noise.}

\label{tab:obs_results}%

\begin{tabular}{lcccccccc}
 \hline
Line& $S_{\text{peak}}$ & FWHM & Velocity & $\rm E_{stas}$  & $\rm E_{sys}$ & $z$& $\Delta$v&$\sigma_{\text{RMS}}$ \\
  &  $\rm mJy\,beam^{-1}$&$\rm km\,s^{-1}$   & $\rm km\,s^{-1}$&$\rm km\,s^{-1}$  & $\rm km\,s^{-1}$ & & $\rm km\,s^{-1}$ & $\rm mJy\,beam^{-1}$ \\
 \hline
H{\sc i}-g1    & $-626.7\pm49.3$& $13.7\pm0.8$&-1.36 &  0.16& $4.2\times10^{-7}$  & $0.24670434 \pm6.7\times10^{-7}$ & 2.0& 4.00  \\
H{\sc i}-g2    & $-150.8\pm23.3$& $8.3\pm0.5$&-11.52 & 0.36 &$4.2\times10^{-7}$ & $0.2466621\pm1.5\times10^{-6}$& 2.0& 4.00  \\
H{\sc i}-g3    & $-77.5\pm41.5$& $11.4\pm2.5$&10.99 &2.66  &$4.2\times10^{-7}$ & $0.246756\pm1.1\times10^{-5}$& 2.0& 4.00 \\
H{\sc i}-g4    & $-27.6\pm1.1$&$107.3\pm3.7$&27.78 & 1.88 & $4.2\times10^{-7}$ & $0.2468255\pm7.8\times10^{-6}$ & 2.0& 4.00 \\
OH 1612    & $-7.0\pm0.1$&$9.7\pm0.2$&-11.14 &0.12  & $2.8\times10^{-3}$ & $0.24666367\pm5.0\times10^{-7}$ &1.8  & 0.29\\
OH 1665    & $-1.1\pm0.1$ &$13.7\pm0.8$&-3.52& 1.03 &  $2.2\times10^{-3}$ & $0.2466953\pm4.3\times10^{-6}$&6.5  & 0.16  \\
OH 1667    & $-0.6\pm0.1$ &$13.7\pm0.8$&-3.56&1.44 &  $7.2\times10^{-4}$ & $0.2466952\pm6.0\times10^{-6}$&6.5  & 0.13  \\
OH 1720    & $7.0\pm0.1$ &$9.7\pm0.2$&-11.19&0.15& $1.7\times10^{-3}$  &$0.24666346\pm6.2\times10^{-7}$&1.7 & 0.36  \\

 \hline
\end{tabular}

\end{sidewaystable}
%\end{table}

\section{Systematic~errors}
Any uncertainties and potential errors would influence the results and thus should be identified. The uncertainties can be classified as statistical and systematic. The former has been determined from the Monte Carlo simulation. The latter cloud arise from a catalog including Earth's motion, frequency calibration, errors in the laboratory frequencies, radio frequency interference (RFI), and astronomical errors arising in the target source.

The data were recorded in topocentric frequency without correction for the Earth's motion during the observations, converting the observed frequencies into the barycentric frame in the data reduction. However, we claim that this does not influence our results at all, because \mbox{H\,{\sc i}} and OH lines were observed simultaneously. This ensures that any velocity shifts due to observational conditions apply equally to all lines, preserving the relative velocity difference between lines which is the key element in our analysis.

Being phase-locked to a maser radio spectrometers are very stable. The relevant uncertainties are ignorable compared to those caused by the errors of laboratory frequencies. Taking the OH 1665 MHz line as an example, its laboratory frequency is 1665.401803 MHz with an error of 0.000012 MHz. This error can cause an uncertainty of $2.2\times10^{-3}~\rm km~s^{-1}$ in the velocity of OH 1665 MHz line. The systematic uncertainties caused by the laboratory frequencies are listed in Table \ref{tab:obs_results}.

Regarding to RFI, all of the OH lines were completely free of this with a minority of the \mbox{H\,{\sc i}} spectra being mildly affected, which have been removed. 

Of the issues arising from source itself, possible effects are the motion of the absorbing gas and variability in the background radio source \citep{combes2023}. However, these variations occur on timescales of years and should not be an issue in the two-hour observation, no matter to mention that the simultaneity of the spectral line observations can ensure that the possible slight uncertainties caused by source itself on the lines can be almost balanced out.

Therefore, the main source of systematic uncertainties is laboratory frequencies in virtue of the simultaneous spectral line observations that canceled out most systematic uncertainties. In comparison, see Table \ref{tab:obs_results}, the systematic uncertainties are ignorable compared to the statistical uncertainties.

\section{Evolution~of~fundamental~constants}
The line frequencies of \mbox{H\,{\sc i}} 21-cm hyperfine transition and OH 18-cm transitions have the dependencies \citep{chengalur2003,curran2004}:

\begin{equation}\label{equ:hi_dpd}
\nu_{21}\propto g_p\mu^{-1}\alpha^{2}R_{\infty},
\end{equation}
\begin{equation}\label{equ:oh_main_dpd_plus}
\nu_{1667}+\nu_{1665}\propto \mu^{-2.57}\alpha^{-1.14}R_{\infty},
\end{equation}
\begin{equation}\label{equ:oh_main_dpd_minus}
\nu_{1667}-\nu_{1665}\propto \mu^{-2.44}\alpha^{-0.88}g_p R_{\infty},
\end{equation}
\begin{equation}\label{equ:oh_sate_dpd_plus}
\nu_{1720}-\nu_{1612}\propto \mu^{-0.72}\alpha^{2.56}g_p R_{\infty},
\end{equation}
where $R_{\infty}\equiv m_ee^{4}/(\hbar^{3}c)$ is the Rydberg constant. Therefore, the line frequency ratios can provide us the variation of $\alpha$, $\mu$, and $g_p$. From \citep{chengalur2003}, the comparison of redshifts:

\begin{itemize}
\item[--] Between $\nu_{1667}+\nu_{1665}$ and  $\nu_{1667}-\nu_{1665}$ gives
\begin{equation}\label{equ:compa_oh_main}
\frac{\Delta z_{1}}{1+z_{1}} = 0.13\frac{\Delta \mu}{\mu}+0.26\frac{\Delta \alpha}{\alpha}+\frac{\Delta g_p}{g_p}=\frac{\Delta (\mu^{0.13}\alpha^{0.26}g_p)}{\mu^{0.13}\alpha^{0.26}g_p},
\end{equation}
where $\Delta z_{1} = z(\nu_{1667}+\nu_{1665})-z(\nu_{1667}-\nu_{1665})$ is the difference in the  measured $\nu_{1667}+\nu_{1665}$ and  $\nu_{1667}-\nu_{1665}$ redshifts  and  $z_{1} = \frac{1}{2}\left[z(\nu_{1667}+\nu_{1665})+z(\nu_{1667}-\nu_{1665})\right]$ is the average.
\\

\item[--] Comparison between $\nu_{1667}+\nu_{1665}$ and H{\sc i} gives
\begin{equation}\label{equ:compa_oh_main_hi}
\frac{\Delta z_{2}}{1+z_{2}} = 1.57\frac{\Delta \mu}{\mu}+3.14\frac{\Delta \alpha}{\alpha}+\frac{\Delta g_p}{g_p}=\frac{\Delta (\mu^{1.57}\alpha^{3.14}g_p)}{\mu^{1.57}\alpha^{3.14}g_p},
\end{equation}
where $\Delta z_{2} = z(\nu_{1667}+\nu_{1665})-z$(H{\sc i}),  $z_{2} = \frac{1}{2}\left[z(\nu_{1667}+\nu_{1665})+z(\text{H{\sc i}})\right]$. 
\\

\item[--] Between $\nu_{1667}-\nu_{1665}$ and H{\sc i} gives
\begin{equation}\label{equ:compa_oh_main2_hi}
\frac{\Delta z_{3}}{1+z_{3}} = 1.44\frac{\Delta \mu}{\mu}+2.88\frac{\Delta \alpha}{\alpha}=\frac{\Delta (\mu^{1.44}\alpha^{2.88})}{\mu^{1.44}\alpha^{2.88}},
\end{equation}
where $\Delta z_{3} = z(\nu_{1667}-\nu_{1665})-z$(H{\sc i}) and  $z_{3} = \frac{1}{2}[z(\nu_{1667}-\nu_{1665})+z$(H{\sc i})].
\\

\item[--] Between $\nu_{1720}+\nu_{1612}$ and $\nu_{1720}-\nu_{1612}$ gives
\begin{equation}\label{equ:compa_oh_sate}
\frac{\Delta z_{4}}{1+z_{4}} = 1.85\frac{\Delta \mu}{\mu}+3.70\frac{\Delta \alpha}{\alpha}+\frac{\Delta g_p}{g_p}=\frac{\Delta (\mu^{1.85}\alpha^{3.70}g_p)}{\mu^{1.85}\alpha^{3.70}g_p},
\end{equation}
where $\Delta z_{4} = z(\nu_{1720}+\nu_{1612})-z(\nu_{1720}-\nu_{1612})$ and $z_{4} = \frac{1}{2}\left[z(\nu_{1720}+\nu_{1612})+z(\nu_{1720}-\nu_{1612})\right]$. 

\end{itemize}

Applying these Equations to our data, we obtained the following results. The comparisons between $\nu_{1667}+\nu_{1665}$ and $\nu_{1667}-\nu_{1665}$, between $\nu_{1667}+\nu_{1665}$ and \mbox{H\,{\sc i}}-g1, between $\nu_{1667}-\nu_{1665}$ and \mbox{H\,{\sc i}}-g1 separately, give:
%%%
\begin{equation}\label{equ:red_set1}
 \rm Set\,1\left\{
  \begin{array}{ll}
   \frac{\Delta \mu}{\mu}+2\frac{\Delta \alpha}{\alpha}+7.75\frac{\Delta g_p}{g_p}=(0\pm4.0)\times10^{-2}&  \text{(a)}  \\
  \frac{\Delta \mu}{\mu}+2\frac{\Delta \alpha}{\alpha}+0.64\frac{\Delta g_p}{g_p}=(-4.7\pm1.9)\times10^{-6}& \text{(b)} \\
   \frac{\Delta \mu}{\mu}+2\frac{\Delta \alpha}{\alpha}=(0\pm3.5)\times10^{-3}& \text{(c).}
  \end{array}
  \right.
\end{equation}
%%%

From the conjugate OH satellite lines we determine $[\Delta Y/Y]=(1.5\pm5.2)\times10^{-6}$. Moreover, we obtain $[\Delta Y/Y]= (-0.9\pm1.3)\times10^{-6}$, by combining it with $[\Delta Y/Y]=(-1.0\pm1.3)\times10^{-6}$ from previous observations \citep{kanekar2018}, weighted based on their uncertainties. Note that $g_{p}$ was implicitly assumed to be non-variable \cite{kanekar2018}, which we do not assume here.

Connecting Equation \ref{equ:red_set1}(b) to $[\Delta Y/Y]= (-0.9\pm1.3)\times10^{-6}$, we further have $\Delta g_p/g_p = (-4.3\pm2.5)\times10^{-5} $ and $\Delta (\mu\alpha^{2})/(\mu\alpha^{2}) =  (2.0\pm1.2)\times10^{-5} $.

\section{Summary}--From the FAST observations of \mbox{H\,{\sc i}} and OH spectral lines toward PKS 1413+135, a stringent constraint, $\Delta g_{p}/g_{p}=(-4.3\pm2.5)\times10^{-5} $, is placed on the cosmological variation of the proton g-factor over a look-back time of 2.9 Gyr. In addition, we obtained $\Delta (\mu\alpha^{2})/(\mu\alpha^{2}) =  (2.0\pm1.2)\times10^{-5}$ and $\Delta (\mu\alpha^{2}g_{p}^{0.64})/(\mu\alpha^{2}g_{p}^{0.64}) = (-4.7\pm1.9)\times10^{-6} $. These are consistent with no evolution. Compared to previous constraints on $g_{p}$ \citep{chengalur2003,su2025,su2025_2}, our result improves the constraint by more than two orders of magnitude. With the FAST, and the future Square Kilometre Array (SKA), we expect further high quality detections \citep{curran2004,chen2019}, leading to experimental constraints on current grand unified theories of physics. 

\section{Acknowledgements}
This work is supported by the National Natural Science Foundation of China (Grant No. 12588202). RZS acknowledges the support from the China Postdoctoral Science Foundation (Grant No. 2024M752979). MFG is supported by the National Science Foundation of China (grant 12473019), the Shanghai Pilot Program for Basic Research-Chinese Academy of Science, Shanghai Branch (JCYJ-SHFY-2021-013), the National SKA Program of China (Grant No. 2022SKA0120102), and the China Manned Space Project with No. CMS-CSST-2025-A07.

This work made use of the data from FAST (Five-hundred-meter Aperture Spherical radio Telescope). FAST is a Chinese national mega-science facility, operated by National Astronomical Observatories, Chinese Academy of Sciences.

\section{Conflict of interest} 
The authors declare no competing interests.

\section{Data availability} 
The raw data has been released by the FAST. The final spectra are available here (https://github.com/SURENZHI/FAST-HI-and-OH-spectra-of-PKS-1413-135).

%%%%%%%%%%%%%%%%%%%%%%%%%%%%%%%%%%%%%%%%%%%%%%%%
%%% Reference section. 
%%% citation in the content using "some words~\cite{1,2}".
%%% ~ is needed to make the reference number is on the same line with the word before it.
%%%%%%%%%%%%%%%%%%%%%%%%%%%%%%%%%%%%%%%%%%%%%%%%%%%%%%%
\clearpage
\bibliographystyle{scpma-zycai} %Citation order, maximum 50 authors
\bibliography{constrain_gp}

\end{document}